\pdfoutput=1
\documentclass[12pt,a4paper]{article}

\usepackage{ifthen} 
\newboolean{pdflatex}
\setboolean{pdflatex}{true} 

\newboolean{articletitles}
\setboolean{articletitles}{true} 

\newboolean{uprightparticles}
\setboolean{uprightparticles}{false} 

\def\paperauthors{Ahmed Abdelmotteleb$^{1}$, Alessandro Bertolin$^{2}$, Chris Burr$^{3}$, Ben Couturier$^{3}$, Ellinor Eckstein$^{4}$, Davide Fazzini$^{5}$, Nathan Grieser$^{6}$, Christophe Haen$^{3}$, Ryunosuke O'Neil$^{3}$, Eduardo Rodrigues$^{7}$, Nicole Skidmore$^{1}$, Mark Smith$^{8}$, Aidan R. Wiederhold$^{9}$, Shunan Zhang$^{10}$}
\def\paperasciititle{The LHCb Sprucing and Analysis Productions} 
\def\papertitle{The \lhcb Sprucing and Analysis Productions} 
\def\paperkeywords{{High Energy Physics}, {LHCb}} 
\def\papercopyright{\the\year\ CERN for the benefit of the LHCb collaboration} 
\def\paperlicence{CC BY 4.0 licence}
\def\paperlicenceurl{https://creativecommons.org/licenses/by/4.0/}

\usepackage[top=1in, bottom=1.25in, left=1in, right=1in]{geometry}

\usepackage{microtype}
\usepackage{lineno}  
\usepackage{xspace} 
\usepackage{caption} 

\usepackage{graphicx}  
\usepackage{color}
\usepackage{colortbl}
\graphicspath{{./figs/}} 

\usepackage{amsmath} 
\usepackage{amssymb}
\usepackage{amsfonts}
\usepackage{upgreek} 

\newcommand*\patchAmsMathEnvironmentForLineno[1]{%
\expandafter\let\csname old#1\expandafter\endcsname\csname #1\endcsname
\expandafter\let\csname oldend#1\expandafter\endcsname\csname
end#1\endcsname
 \renewenvironment{#1}%
   {\linenomath\csname old#1\endcsname}%
   {\csname oldend#1\endcsname\endlinenomath}%
}
\newcommand*\patchBothAmsMathEnvironmentsForLineno[1]{%
  \patchAmsMathEnvironmentForLineno{#1}%
  \patchAmsMathEnvironmentForLineno{#1*}%
}
\AtBeginDocument{%
\patchBothAmsMathEnvironmentsForLineno{equation}%
\patchBothAmsMathEnvironmentsForLineno{align}%
\patchBothAmsMathEnvironmentsForLineno{flalign}%
\patchBothAmsMathEnvironmentsForLineno{alignat}%
\patchBothAmsMathEnvironmentsForLineno{gather}%
\patchBothAmsMathEnvironmentsForLineno{multline}%
\patchBothAmsMathEnvironmentsForLineno{eqnarray}%
}

\usepackage[pdftex,
            pdfauthor={\paperauthors},
            pdftitle={\paperasciititle},
            pdfkeywords={\paperkeywords},
            ]{hyperref}

\usepackage[colorinlistoftodos,textsize=scriptsize]{todonotes}

\usepackage[bottom,flushmargin,hang,multiple]{footmisc}

\usepackage[all]{hypcap} 

\usepackage{xspace} 
\usepackage{upgreek}

\def\lhcb   {\mbox{LHCb}\xspace}

\def\lhc    {\mbox{LHC}\xspace}

\def\MagUp {\mbox{\em Mag\kern -0.05em Up}\xspace}

\def\hlttwo {HLT2\xspace}

\ifthenelse{\boolean{uprightparticles}}%
{

 \def\PDelta      {\ensuremath{\Delta}\xspace}                 
 \def\PXi         {\ensuremath{\Xi}\xspace}                 
 \def\PLambda     {\ensuremath{\Lambda}\xspace}                 
 \def\PSigma      {\ensuremath{\Sigma}\xspace}                 
 \def\POmega      {\ensuremath{\Omega}\xspace}                 
 \def\PUpsilon    {\ensuremath{\Upsilon}\xspace}
 \let\oldPi\Pi
 \def\PPi         {\ensuremath{\oldPi}\xspace}

 \def\PB      {\ensuremath{\mathrm{B}}\xspace}                 
                  
 \def\PD      {\ensuremath{\mathrm{D}}\xspace}

 \def\PK      {\ensuremath{\mathrm{K}}\xspace}

 \def\Pb      {\ensuremath{\mathrm{b}}\xspace}

 \def\Pi      {\ensuremath{\mathrm{i}}\xspace}

 \def\Ps      {\ensuremath{\mathrm{s}}\xspace}

 \def\thebaroffset{0.0em}
}
{

 \mathchardef\PDelta="7101
 \mathchardef\PXi="7104
 \mathchardef\PLambda="7103
 \mathchardef\PSigma="7106
 \mathchardef\POmega="710A
 \mathchardef\PUpsilon="7107
 \mathchardef\PPi="7105
                  
 \def\PB      {\ensuremath{B}\xspace}                 
                  
 \def\PD      {\ensuremath{D}\xspace}

 \def\PK      {\ensuremath{K}\xspace}

 \def\Pb      {\ensuremath{b}\xspace}

 \def\Pi      {\ensuremath{i}\xspace}

 \def\Ps      {\ensuremath{s}\xspace}

 \def\thebaroffset{0.18em}
}
\newcommand{\offsetoverline}[2][\thebaroffset]{\kern #1\overline{\kern -#1 #2}}%

\makeatletter
\ifcase \@ptsize \relax
  \newcommand{\miniscule}{\@setfontsize\miniscule{4}{5}}
\or
  \newcommand{\miniscule}{\@setfontsize\miniscule{5}{6}}
\or
  \newcommand{\miniscule}{\@setfontsize\miniscule{5}{6}}
\fi
\makeatother

\DeclareRobustCommand{\optbar}[1]{\shortstack{{\miniscule (\rule[.5ex]{1.25em}{.18mm})}
  \\ [-.7ex] $#1$}}

\def\squark    {{\ensuremath{\Ps}}\xspace}

\def\bquark    {{\ensuremath{\Pb}}\xspace}
\def\bquarkbar {{\ensuremath{\overline \bquark}}\xspace}

\def\KorKbar {\kern \thebaroffset\optbar{\kern -\thebaroffset \PK}{}\xspace}

\def\D       {{\ensuremath{\PD}}\xspace}

\def\DorDbar {\kern \thebaroffset\optbar{\kern -\thebaroffset \PD}\xspace}
\def\Dz      {{\ensuremath{\D^0}}\xspace}

\def\Dp      {{\ensuremath{\D^+}}\xspace}
\def\Dm      {{\ensuremath{\D^-}}\xspace}

\def\DpDm    {\ensuremath{\Dp {\kern -0.16em \Dm}}\xspace}

\def\Dstarp  {{\ensuremath{\D^{*+}}}\xspace}

\def\B       {{\ensuremath{\PB}}\xspace}

\def\BorBbar {\kern \thebaroffset\optbar{\kern -\thebaroffset \PB}\xspace}

\def\Bd      {{\ensuremath{\B^0}}\xspace}

\def\BdorBdbar {\kern \thebaroffset\optbar{\kern -\thebaroffset \Bd}\xspace}

\def\Bs      {{\ensuremath{\B^0_\squark}}\xspace}

\def\BsorBsbar {\kern \thebaroffset\optbar{\kern -\thebaroffset \Bs}\xspace}

\def\Y#1S{\ensuremath{\PUpsilon{(#1S)}}\xspace}

\def\LorLbar     {\kern \thebaroffset\optbar{\kern -\thebaroffset \PLambda}\xspace}

\def\order   {{\ensuremath{\mathcal{O}}}\xspace}

\def\AT#1     {\ensuremath{A_{\mathrm{T}}^{#1}}\xspace}           

\def\C#1      {\ensuremath{\mathcal{C}_{#1}}\xspace}                       
\def\Cp#1     {\ensuremath{\mathcal{C}_{#1}^{'}}\xspace}                    
\def\Ceff#1   {\ensuremath{\mathcal{C}_{#1}^{\mathrm{(eff)}}}\xspace}        
\def\Cpeff#1  {\ensuremath{\mathcal{C}_{#1}^{'\mathrm{(eff)}}}\xspace}       
\def\Ope#1    {\ensuremath{\mathcal{O}_{#1}}\xspace}                       
\def\Opep#1   {\ensuremath{\mathcal{O}_{#1}^{'}}\xspace}                    

\newcommand{\aunit}[1]{\ensuremath{\text{\,#1}}}       

\newcommand{\tev}{\aunit{Te\kern -0.1em V}\xspace}
\newcommand{\gev}{\aunit{Ge\kern -0.1em V}\xspace}
\newcommand{\mev}{\aunit{Me\kern -0.1em V}\xspace}
\newcommand{\kev}{\aunit{ke\kern -0.1em V}\xspace}
\newcommand{\ev}{\aunit{e\kern -0.1em V}\xspace}
 
\newcommand{\mevc}{\ensuremath{\aunit{Me\kern -0.1em V\!/}c}\xspace}
\newcommand{\gevc}{\ensuremath{\aunit{Ge\kern -0.1em V\!/}c}\xspace}
\newcommand{\mevcc}{\ensuremath{\aunit{Me\kern -0.1em V\!/}c^2}\xspace}
\newcommand{\gevcc}{\ensuremath{\aunit{Ge\kern -0.1em V\!/}c^2}\xspace}

\def\fb   {\ensuremath{\aunit{fb}}\xspace}
\def\invfb   {\ensuremath{\fb^{-1}}\xspace}

\def\order{{\ensuremath{\mathcal{O}}}\xspace}

\def\gsim{{~\raise.15em\hbox{$>$}\kern-.85em
          \lower.35em\hbox{$\sim$}~}\xspace}
\def\lsim{{~\raise.15em\hbox{$<$}\kern-.85em
          \lower.35em\hbox{$\sim$}~}\xspace}

\def\davinci    {\mbox{\textsc{DaVinci}}\xspace}
\def\decaytreefitter {\mbox{\textsc{DecayTreeFitter}}\xspace}
\def\dirac      {\mbox{\textsc{Dirac}}\xspace}
\def\gaudi      {\mbox{\textsc{Gaudi}}\xspace}

\def\moore      {\mbox{\textsc{Moore}}\xspace}

\def\root       {\mbox{\textsc{Root}}\xspace}

\def\gbyps      {\aunit{GB/s}\xspace}

\def\pbytes     {\aunit{PB}\xspace}

\def\tell1  {TELL1\xspace}
\def\ukl1   {UKL1\xspace}

\newcommand{\eg}{\mbox{\itshape e.g.}\xspace}
\newcommand{\ie}{\mbox{\itshape i.e.}\xspace}

\newcommand{\lhcborcid}[1]{\href{https://orcid.org/#1}{\hspace*{0.1em}\raisebox{-0.45ex}{\includegraphics[width=1em]{figs/orcidIcon.pdf}}}}

\usepackage{cite} 
\usepackage{mciteplus}

\usepackage{longtable} 
\usepackage{tabularx}
\usepackage{booktabs}
\usepackage{tikz}
\usepackage{graphicx}
\usepackage{subcaption}
\usepackage{hyperref}
\usepackage{fontawesome}
\usepackage{rotating}
\usepackage{booktabs}
\usepackage{color,soul}

\begin{document}

\renewcommand{\thefootnote}{\fnsymbol{footnote}}
\setcounter{footnote}{1}


\begin{titlepage}
\pagenumbering{roman}

\noindent
\begin{tabular*}{\linewidth}{lc@{\extracolsep{\fill}}r@{\extracolsep{0pt}}}

 & & \today \\ 
\hline
\end{tabular*}

\vspace*{2.0cm}

{\normalfont\bfseries\boldmath\huge
\begin{center}
  \papertitle 
\end{center}
}

\vspace*{1.0cm}

\begin{center}
\paperauthors.
\bigskip\\
{\normalfont\itshape\footnotesize
$ ^1$Department of Physics, University of Warwick, Coventry, United Kingdom, \\
$ ^2$INFN Sezione di Padova, Padova, Italy, \\
$ ^3$European Organization for Nuclear Research (CERN), Geneva, Switzerland, \\
$ ^4$Universit\"{a}t Bonn - Helmholtz-Institut f\"{u}r Strahlen und Kernphysik, Bonn, Germany,\\
$ ^{5}$INFN Sezione di Milano-Bicocca, Milano, Italy,
$ ^6$University of Cincinnati, Cincinnati, USA,\\
$ ^7$Oliver Lodge Laboratory, University of Liverpool, Liverpool, United Kingdom,\\
$ ^8$Imperial College London, London, United Kingdom,\\
$ ^9$Department of Physics and Astronomy, University of Manchester, Manchester, United Kingdom, \\
$ ^{10}$Department of Physics, University of Oxford, Oxford, United Kingdom, \\

}
\end{center}

\vspace{\fill}

\begin{abstract}
\noindent 

The \lhcb detector underwent a comprehensive upgrade in preparation for the third data-taking run of the Large Hadron Collider (LHC), known as \lhcb Upgrade I. With its increased data rate, Run 3 introduced considerable challenges in both data acquisition (online) and data processing and analysis (offline). The offline processing and analysis model was upgraded to handle the factor 30 increase in data volume and the associated demands of ever-growing datasets for analysis, led by the \lhcb Data Processing and Analysis (DPA) project. This paper documents the \lhcb ``Sprucing" --- the centralised offline processing, selections and streaming of data --- and ``Analysis Productions" --- the centralised and highly automated declarative nTuple production system. The \davinci application used by analysis productions for tupling spruced data is described as well as the {\tt apd} and {\tt lbconda} tools for data retrieval and analysis environment configuration. These tools allow for greatly improved analyst workflows and analysis preservation. Finally, the approach to data processing and analysis in the High-Luminosity Large Hadron Collider (HL-LHC) era --- \lhcb Upgrade II --- is discussed.

\end{abstract}

\vspace*{2.0cm}

\vspace{\fill}

{\footnotesize 
\centerline{\copyright~\papercopyright. \href{\paperlicenceurl}{\paperlicence}.}}
\vspace*{2mm}

\end{titlepage}


\newpage
\setcounter{page}{2}
\mbox{~}
%
%
%
%


\renewcommand{\thefootnote}{\arabic{footnote}}
\setcounter{footnote}{0}

\cleardoublepage


\pagestyle{plain} 
\setcounter{page}{1}
\pagenumbering{arabic}


\tableofcontents

\newpage
\section{Introduction}

The \lhcb experiment is one of the four main experiments collecting data from proton-proton collisions at the Large Hadron Collider (LHC)~\cite{Alves:2008zz}. It is a forward arm spectrometer specialising in the decays of beauty and charm hadrons. During the first and second LHC data collection runs --- Runs~1~\&~2 --- \lhcb collected data corresponding to an integrated luminosity of 9\invfb, equating to over $10^{12}$ \bquark \bquarkbar pairs in the acceptance of the detector.
In 2022 the LHC commenced its third run of data-taking known as Run 3. For Run 3 \lhcb underwent a comprehensive upgrade --- known as \lhcb Upgrade I --- in anticipation of a factor 5 increase in delivered luminosity~\cite{LHCb:upgradeone}. This equated to an increase of more than a factor 30 in the volume of data collected by \lhcb per unit of time, taking into account the increase in delivered instantaneous luminosity, a factor 3 due to the increased average event size, and a factor 2 due to higher trigger efficiencies~\cite{LHCbCollaboration:2319756}.
Run 3 therefore posed not only data collection (online) challenges, but also significant offline data processing and analysis ones. This paper documents the two key developments to facing these challenges: 

\begin{description}
\item \textbf{Sprucing}: Centralised offline data skimming, slimming and streaming to reduce the data footprint between tape and disk storage.
\item \textbf{Analysis productions}: Centralised {\tt nTuples} production using the \lhcb{\dirac}~\cite{Tsaregorodtsev:2010zz} transformation system with maximal automation and optimised user experience.
\end{description}

\begin{figure}
    \centering   \includegraphics[width=\textwidth]{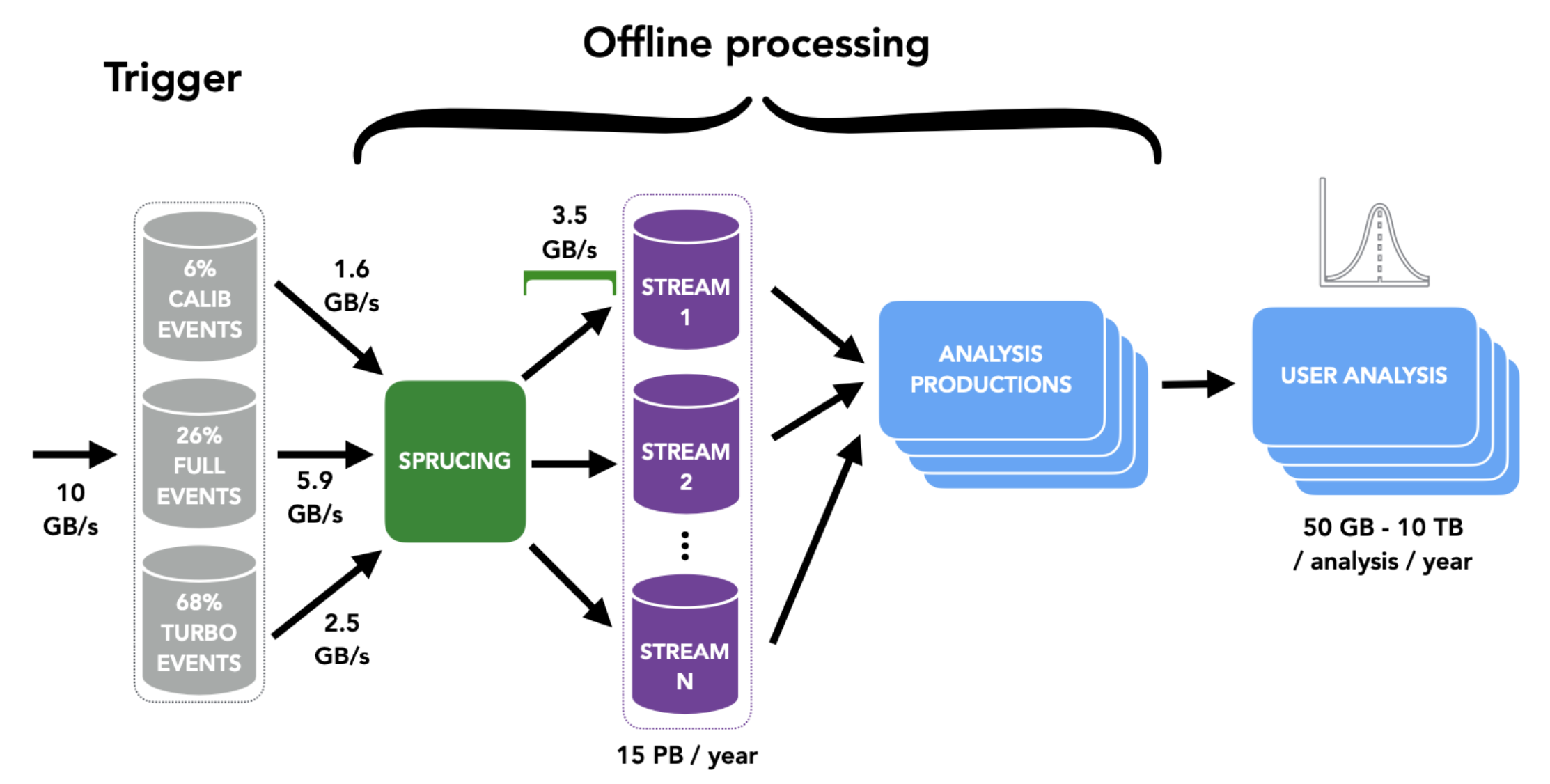}
    \caption{Schematic of the \lhcb offline data flow. Figure adapted from Ref.~\cite{LHCb-FIGURE-2020-016}.}
    \label{fig:DPA:dataflow}
\end{figure}

\noindent The \lhcb Upgrade I offline data and processing flow is sketched in Figure~\ref{fig:DPA:dataflow}, showing the role of the Sprucing and Analysis Productions.

\section{Sprucing}
\label{sec:wp1}

\subsection{The Role of the Sprucing in \lhcb Upgrade I}
\label{sec:roleofsprucing}

\begin{figure}
    \centering
    \includegraphics[width=\textwidth]{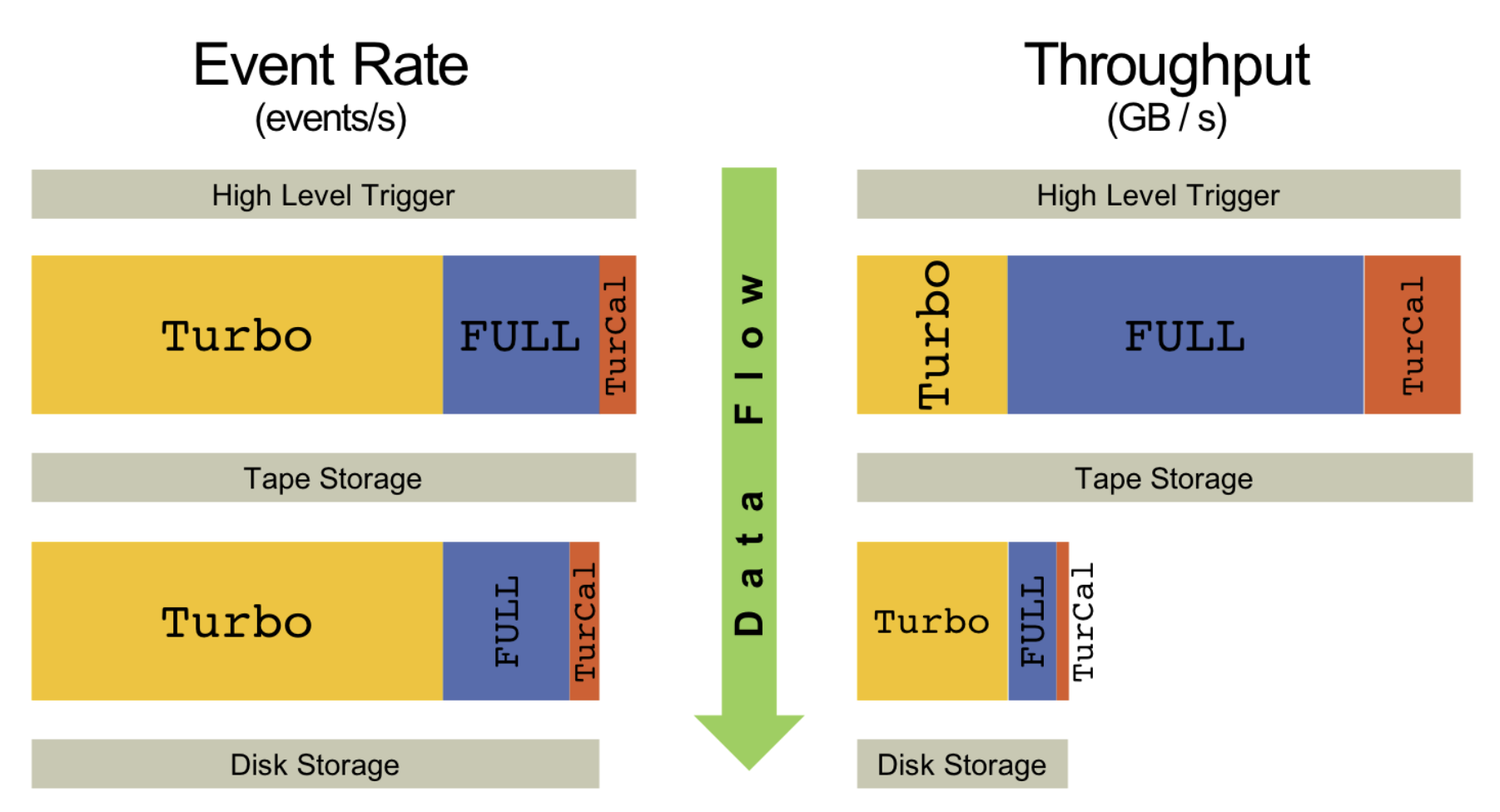}
    \caption{The Sprucing reductions in event rate and throughput for the {\tt FULL}, {\tt TURBO} and {\tt TURCAL} streams. Left: event rates. Right: throughput in \gbyps. Box widths are proportional to the corresponding quantities. Figure taken from Ref.~\cite{LHCbCollaboration:2319756}.}
    \label{fig:Sprucing_BW}
\end{figure}

The LHCb High-Level Trigger (HLT) comprises two stages --- HLT1 and HLT2. The \lhcb online system writes events that pass the HLT2 to tape in three physics streams, discussed below: {\tt FULL}, {\tt TURBO}, and {\tt TURCAL} \footnote{There are an additional three technical streams that are written.}. A physics event will populate one of these streams if it passes a HLT2 selection line that belongs to the stream. In Run~3 \lhcb has a design data rate to tape of 10\gbyps. The primary purpose of the Sprucing is to reduce this data rate to 3.5\gbyps where it can reasonably be saved to disk and therefore be accessible to analysts. How the Sprucing achieves this rate reduction is stream dependent, as described in the following section; Figure~\ref{fig:Sprucing_BW} illustrates the methods used for each physics stream.

The \lhcb online processing produces and exports data in the Mast Data Files {\tt MDF} format~\cite{mdf}. In {\tt MDF} file format, each event is stored sequentially such that files can be merged by concatenation. Problematic files can be partially recovered by starting the processing from the first valid event. Although the {\tt MDF} format provides many convenient features for the \lhcb data acquisition system, it is limited to only storing per-event data. Furthermore, {\tt MDF} files tend to be large in size due to the lack of inter-event compression. To mitigate the effect of large file sizes, {\tt MDF} files are compressed with {\tt Zstandard}~\cite{rfc} when exported from the \lhcb detector site to WLCG resources.

The {\tt MDF} files exported from \lhcb contain so-called \textit{RawEvents}, each of which consist of a collection of \textit{RawBanks} containing different types of event data. A RawBank can store:

\begin{itemize}
    \item  Physics candidate information such as the 4-momenta of particles in a decay-chain and any reconstruction objects requested. This is known as the ``DstData" RawBank.
    \item HLT\{1,2\} or Sprucing line decisions. These are called the ``HltDecReports" RawBanks.
    \item Detector response information - eg. from the calorimeter system ({\tt CALO}) or muon chambers ({\tt MUON}).
\end{itemize}

The Sprucing step outputs so-called {\tt DST} files in the {\tt ROOT} file format, which again consist of, but are not limited to, RawEvents consisting of RawBanks. These {\tt DST} files are distributed to World-wide LHC Computing Grid (WLCG)~\cite{bird2011wlcg} sites and saved on disk to be available to analysts.


\subsubsection*{The {\tt FULL} stream}

The {\tt FULL} stream contains inclusive HLT2 selection lines such as the {\tt topological} lines, and the full event reconstruction persists in perpetuity. This also enables \lhcb's legacy physics programme, exploiting the \lhcb dataset for new physics channels long after \lhcb stops taking data.

For the {\tt FULL} stream the Sprucing slims and skims events by running additional, exclusive selection lines on top of these inclusive events offline. The Sprucing runs selections using the same application --- the \moore application~\cite{Moore} --- as the online system, and hence HLT2 selection lines and Sprucing selection lines are identical and, by design, trivially interchangeable.  Furthermore, the same algorithms and tools are shared between HLT2, Sprucing and the offline analysis software project \davinci~\cite{DaVinci}, namely the {\sc ThOr}~\cite{Thor} based selection and combinatorial algorithms. This was a conscious choice and departure from the Run~2 model. Compared to HLT2, the Sprucing selection lines benefit from less strict limits on the timing of selection algorithms. This makes the {\tt FULL} stream particularly important for physics channels with many final-state particles, which may suffer from large combinatorial backgrounds. Lines such as $B^0 \rightarrow D^{*} h h h$ with $D^{*} \rightarrow D^0 \pi$ and 
$D^0 \rightarrow h h h h$ where $h = K, \pi$, unavoidably require the computation of a large number of particle combinations to build vertices and composite particles that can be selected. The consequent reduction in throughput means that these lines cannot be run at the HLT2 stage and must instead be run by the Sprucing on the data saved by the inclusive lines of the {\tt FULL} stream. 


The Sprucing, via exclusive selection lines, is required to achieve a factor 7 reduction in the {\tt FULL} stream bandwidth so that this data can be saved to disk. For efficient data access by analysts, the Sprucing is tasked with further streaming the data and creates \order{(20)} streamed {\tt DST} outputs based on physics case and line rate. The Sprucing step also creates File Summary Records (FSRs) that record luminosity information useful for offline analysis, meaning that the luminosity events can be discarded at this stage. FSRs are described in more detail in Subsection~\ref{sec:fsrs}. 

The advantage of the {\tt FULL} stream model is that, due to the persistence of the full event reconstruction, this data can be re-processed with new or updated selection lines periodically many years into the future as discussed in Section~\ref{sec:resprucing}.

At the end of the 2024 data taking period, the {\tt FULL} stream contained 401 (mainly) inclusive HLT2 lines, and the Sprucing subsequently ran 1138 Sprucing lines on this stream with the output saved to disk. 

\subsubsection*{The {\tt TURBO} stream}

In the {\tt TURBO} stream~\cite{Aaij:2147693}, HLT2 selection lines are generally exclusive to a particular decay channel and persist only a custom set of physics objects and their reconstruction, minimally the triggering candidate and primary vertices in an event. This became the Run~3 default model, and to achieve the baseline of 10\gbyps to tape, \lhcb required 73\% of its physics programme to use {\tt TURBO}. In general, lines that select well-known simple decay topologies, such as $B^0 \rightarrow D^- \pi^+$, are included in the {\tt TURBO} stream. With relatively simple selections, it is possible to achieve sufficiently small rates, consistent with the overall requirement of a 2.5 \gbyps bandwidth to disk. It should be noted that the {\tt TURBO} model is highly flexible and any object including raw detector banks can be selectively persisted. 
In the {\tt TURBO} stream, events are already exclusively selected for a dedicated decay channel with only the necessary event information persisted for that decay. Therefore, no further slimming of this data is required by the Sprucing. The Sprucing writes the {\tt TURBO} data to \order{(20)} sub-streams and populates the FSRs.
At the end of the 2024 data taking period, the {\tt TURBO} stream was running 2502 lines, which were subsequently spruced.

\subsubsection*{The {\tt TURCAL} stream}

Finally, the {\tt TURCAL} stream contains selection lines for physics channels that are used for, among other studies, detector calibration. Hence these lines additionally persist raw detector information ({\tt CALO} or {\tt MUON} for instance). In the case of {\tt TURCAL}, a factor 8 reduction is again required by the Sprucing and the data is further streamed. The {\tt TURCAL} Sprucing achieves this by facilitating a line-by-line customisation of the persistency of reconstruction objects and detector RawBanks to reduce the data footprint to disk. The {\tt TURCAL} stream, like the {\tt FULL} stream, maintains the option for re-processings and even re-running the reconstruction due to the presence of raw detector information.
At the end of the 2024 data taking period, the {\tt TURCAL} stream was running 147 lines, which were subsequently Spruced.


\subsection{Sprucing Productions and Campaigns}

The Sprucing runs in centrally-managed offline productions using the \lhcb{\dirac} transformation system on WLCG resources. Productions are created through YAML files and tracked through GitLab~\cite{git, gitlab} issues by Sprucing coordinators and production managers. Continuous Integration is used for selection line development --- performed by hundreds of collaboration members --- and maintenance with automated rate and bandwidth tests performed on a nightly basis or triggered on demand~\cite{arXiv:2503.19582}. Whenever possible, these tests use ad-hoc
recorded data samples as input. This ensures a fully tested suite of selection lines at all times and avoids the line
preparation campaigns of Runs~1~\&~2 that lasted several weeks.

\subsubsection*{Concurrent Sprucing Campaigns}

Similarly to Stripping of Runs~1~\&~2~\cite{LHCb-TDR-011}, the Sprucing runs concurrently with data taking. Multiple Sprucing campaigns are established over a data-taking year corresponding to changes in the configuration of HLT1, HLT2 or the Sprucing itself that impacts the physics output, for example, updates to existing selection lines or the addition of new ones. The development schedule of trigger and Sprucing campaigns is closely aligned and coordinated.

Sprucing campaigns are validated efficiently via the use of Analysis Productions (described in detail in Section~\ref{sec:anaprods}). {\tt nTuples} are produced from a small subset of Spruced files that analysts can verify interactively via the Analysis Productions web browser integrated with tools such as {\tt JavaScriptROOT}~\cite{ROOT,jsroot}.

In nominal, concurrent campaigns, the observed turnaround between the HLT2 processing, file transfer to the Offline system, and the Sprucing processing is 24-48 hours.  

\subsubsection*{Re-Sprucing Campaigns}
\label{sec:resprucing}

Re-Sprucing campaigns take place in end-of-year LHC shutdowns when the Sprucing input data can be staged from tape to disk and Offline resources have capacity. This (typically) four-month period constitutes a strict time frame within which the re-Sprucing must be completed to avoid conflict with concurrent campaigns.

Given the full event reconstruction persisted in the {\tt FULL} stream, re-Sprucings offer the opportunity to re-optimise selections or add new selection lines based on the needs of physics analysts. The preparation of re-Sprucing campaigns starts several weeks before the end of data-taking of the year, when analysts submit requests and implement changes to selection lines. 

Once the concurrent data processing has finished, the Sprucing input data is staged from tape to disk. A small fraction of data is re-Spruced as part of the campaign validation with Analysis Productions, again, used by analysts to check distributions of variables. Once the green-light is given, the full production is launched in order to complete the processing before the start of the next year's data-taking.

Re-Sprucing campaigns can either be {\it full} --- whereby all Sprucing lines are rerun and the concurrent campaign can be superseded --- or  {\it incremental} --- where only new or updated lines run and this dataset is additional to the previous {\it full} (re-)Sprucing dataset. The decision between running an {\it incremental} or {\it full} campaign is made based on the number of lines added/modified and the resulting storage requirements.




\subsubsection*{Heavy Ion Data Runs}

\lhcb not only takes event data from proton-proton ($pp$) collisions, but also heavy-ion collisions provided by the LHC. This data is also Spruced. Two physics streams are output from the HLT2: {\tt ION} and {\tt IONRAW}. The {\tt IONRAW} stream undergoes a simple pass through in the Sprucing, similar to $pp$ {\tt TURBO} data. However, the {\tt ION} stream undergoes a similar processing as the $pp$ {\tt TURCAL} stream where, line by line, the persistency of reconstruction objects and detector RawBanks can be customised, reducing the data footprint to disk. This data can be re-Spruced in the future should the extra persistency objects that are on tape be required.

\subsection{File Summary Records (FSRs)}
\label{sec:fsrs}

A File Summary Record (FSR) is a per-file tree data structure containing information on the {\tt DST}  file content, stored within the {\tt DST} file itself. FSRs cannot be used in the \lhcb~{\tt MDF} file format due to their ``per-event" storage as described in Section \ref{sec:roleofsprucing}, so Sprucing is the first point in the data flow in which FSRs can be created. FSRs have been used since \lhcb Run~1 to record luminosity information. The total number of luminosity events in a file, identified by a unique routing bit, is recorded inside the FSR to indicate the \lhcb luminosity a file represents. Storing this information in an FSR means that these events can be discarded at the Sprucing stage.

In Run~3 the FSR information was expanded to include the decoding tables and the application options of the last processing step making Spruced files fully self-contained. 
As described above, \lhcb data contains a ``DstData" RawBank for physics candidate information and a ``HltDecReports" RawBank for the trigger decisions of each event. To save disk space, this data is encoded. The (potentially long) strings of physics object locations ({\tt PackedObjectLocations}) and the selection line decision names ({\tt HLT1SelectionID}, {\tt HLT2SelectionID}, {\tt SpruceSelectionID}) are mapped to integer values. In the RawBanks, only the integers and corresponding information are stored per event and so, in order to read the data, these integer-to-string mappings, referred to as decoding tables, are necessary. Each decoding table is identified by a unique hexadecimal decoding key and stored. To make the Sprucing output files self-contained, the decoding tables are now written to the FSR and can be read from there in further processing steps. 

The application options used to create a Spruced file are also now stored in the file's FSR --- these are the \moore options used to run the Sprucing job. Besides the provenance aspect of knowing exactly how a file was produced via metadata stored in the file itself, storing these options allows further automation of Analysis Productions (described in Section \ref{sec:anaprods}),
as well as configuration consistency checks, as configuration options to run the \davinci application on Spruced files can be deduced from the FSR. These configuration options include the detector geometry version and data-taking conditions, the file type, the input process (\eg, {\tt HLT2}) and whether the file contains real or simulated data.

\subsection{Sprucing Performance in Run~3}

\begin{table}
    \centering
    \begin{tabular}{c | c | c}
        Stream & Stream size before Sprucing (PB) & Stream size after Sprucing (PB)\\
        \hline
        {\tt FULL} & 3.08 & 0.45 \\
        {\tt TURBO} & 1.25 & 1.12 \\
        {\tt TURCAL} & 0.86 & 0.11 \\
    \end{tabular}
    \caption{Stream size before and after Sprucing for data taking in 2024. The stream size after Sprucing is calculated as the sum over all the sub-streams created from the stream by the Sprucing.}
    \label{tab:streamsize}
\end{table}

\begin{table}
    \centering
    \begin{tabular}{c | c | c}
        Stream & Av. evt size before Sprucing (kB) & Av. evt size after Sprucing (kB)\\
        \hline
        {\tt FULL} & 114.9 & 33.8 \\
        {\tt TURBO} & 13.4 & 10.2 \\
        {\tt TURCAL} & 176.9 & 21.2 \\
    \end{tabular}
    \caption{Average event size before and after Sprucing for data taking in 2024.}
    \label{tab:avevtsize}
\end{table}



\begin{figure}
    \centering
    \includegraphics[width=\textwidth]{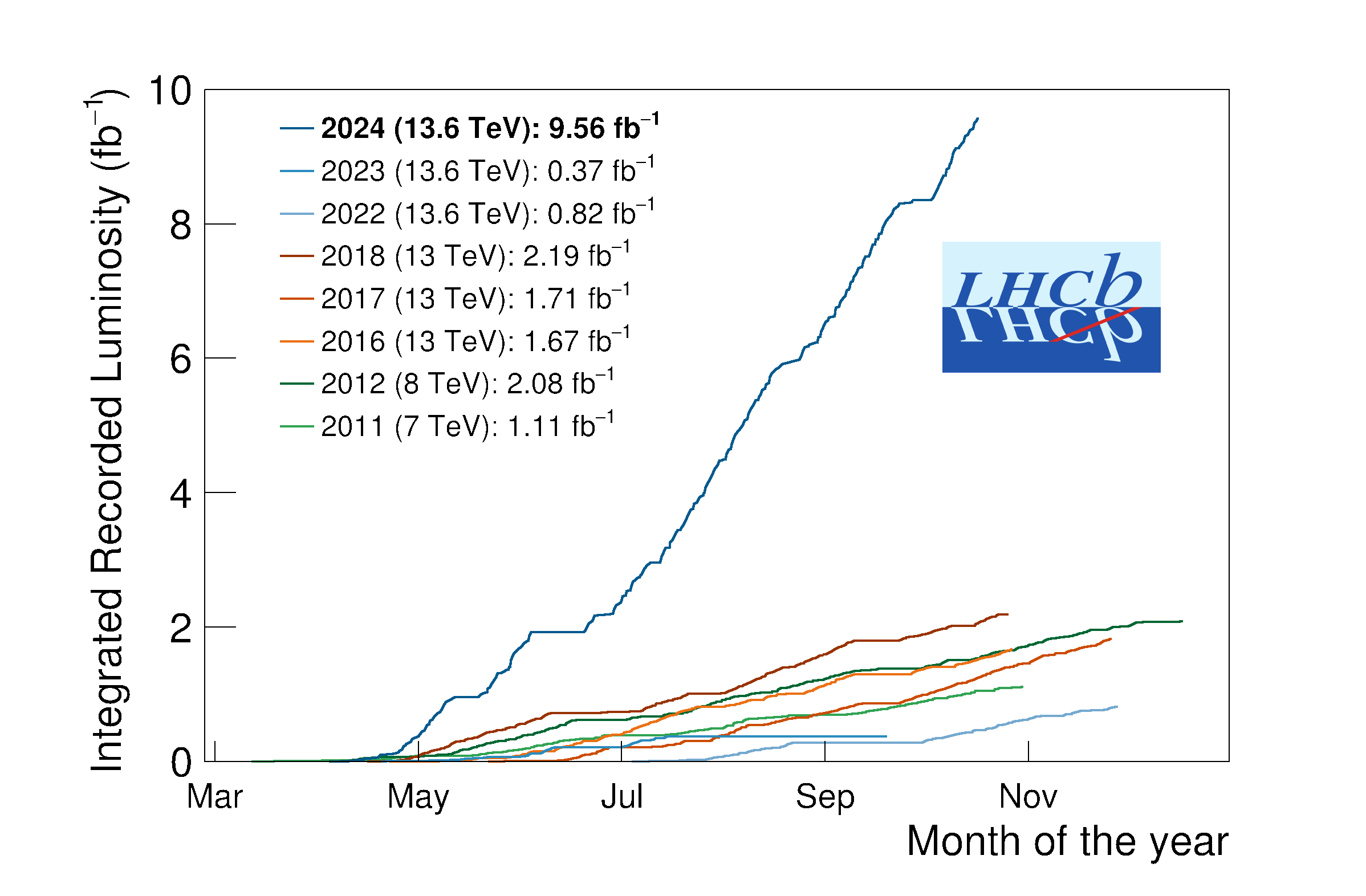}
    \caption{Integrated luminosity recorded by \lhcb per data-taking year from 2011 to 2024. Figure taken from Ref.~\cite{lumiplotcitation}.}
    \label{fig:lumiplot}
\end{figure}

In 2024 \lhcb recorded a record 9.56\invfb of data as shown in Figure~\ref{fig:lumiplot}. Bandwidth reductions of a factor 7 and 8 were achieved for the {\tt FULL} and {\tt TURCAL} streams, respectively. The stream sizes before and after the Sprucing can be seen in Table~\ref{tab:streamsize} with the corresponding average event sizes in Table~\ref{tab:avevtsize}. Due to the efficiency of the concurrent Sprucing productions and subsequent Analysis Productions (as described in section~\ref{sec:anaprods}) analysts were able to begin analysing data within 2--3 days of it being recorded --- a first for any LHC experiment and complementing the online Real-Time Analysis (RTA) ethos. 

\section{Analysis Productions}
\label{sec:anaprods}

\subsection{The Role of Analysis Productions in \lhcb Upgrade I}

After the Sprucing procedure, the data is split into multiple streams (Figure~\ref{fig:DPA:dataflow}), which are distributed to WLCG sites and made directly accessible to analysts. In the Run 1\&2 analysis model, individual analysts would submit user jobs to \lhcb{\dirac} to create {\tt nTuples} from these files. The analyst would have to monitor the jobs, handle issues in the workflow in case of failure, and manage the storing of the output.
Although this pragmatic approach (the ``user job model") was practicable for the Run~1~and/or~2 dataset, the approach does not scale for the following reasons: 

\begin{description}
    \item \textbf{Error recovery:} With larger input data samples comes higher occurrences of faults in data processing or grid operations. In many cases, these require expert attention to recover from, for which the end-user is typically not equipped. Additionally, issues are often common among many analysts. In the ``user-job" model, more work is done to fix these issues as it is not addressed through a single, central solution.

    \item \textbf{Human error:} The cost (in resources and storage) of configuration errors increases with significantly larger input and job volumes.

    \item \textbf{Operational burden:} Problematic user workflows can interfere with the wider operation of the Grid such as centralised data or simulation productions.

    \item \textbf{Data lifecycle:} The lifecycle of the output data is not systematically managed. 
    This data forms the basis of the analysis results and scientific output, and in the user job model the relevant provenance information is highly prone to loss.

\end{description}

Analysis Productions (APs) are an extension of the \lhcb{\dirac} transformation system that enables a simplified and declarative approach to the configuration of distributed computing workflows with \lhcb applications. The majority use case in Run~3 is the transformation of the \lhcb dataset into {\tt nTuples}, rectangular datasets used for physics analysis, which are produced by the \lhcb application \davinci described in more detail in Section~\ref{sec:davinci}. APs provide a simplified user interface to configuring and testing workflows on the WLCG's vast and complex resources and constitute an essential and transformative development for \lhcb in Run~3.

\subsection{Declarative Tupling and the User Interface}

Analysis Productions may be configured to run released \lhcb applications or scripts, where each application or script (a step) transforms the input data into one or more output files. Steps can be chained together to transform the \lhcb data in a variety of ways, but generally the first step for analysts is to run the offline analysis application \davinci to create {\tt nTuples}.


Analysis Productions are defined declaratively via YAML files by describing the workflow, job configurations, and bookkeeping query for the input data, the latter enabling \lhcb{\dirac} to automatically adjust the way in which files are grouped and handle failures.
Information about productions and the provenance of files is permanently stored in the so-called \textit{\lhcb bookkeeping system} catalog~\cite{bkk}. enabling high quality analysis preservation and additional safety checks to be performed, as well as efficient dataset cleanup and archival. 

Analysis Productions are submitted by users via merge requests to a GitLab repository. Minimally, this merge request contains \lhcb application scripts --- in most cases a \davinci script --- and the YAML describing the job. The AP repository is linked with \lhcb{\dirac} through custom services that handle production testing and submission. 

\subsection{Production Testing and Submission}

To validate user-prepared configurations and avoid wasted computing resources and instability in grid operations, extensive continuous integration tests are administered and supported by custom services (an API, celery task queue, and database) linked with \lhcb{\dirac}. Tests are triggered by updates to the merge request. Submission of real productions are gated by the result of these tests and the approval from physics working group liaisons and, in some cases, \lhcb grid experts. The validation process includes: 

\begin{description}
    \item \textbf{Job stability:} Running the entire chain using a small (configurable) subset of input data.
    \item \textbf{Memory footprint:} Detailed analysis of the application memory consumption versus run time using the programme {\tt prmon}~\cite{prmon}.
    \item \textbf{Expected storage footprint:} Estimation of the final {\tt nTuple} sample size extrapolating from the validation results.
\end{description}


Should any monitored criteria exceed predefined thresholds, the tests will fail, and suggestions to fix the problem are provided to the user (via the merge request) on the AP GitLab platform. The productions, their status, and details of the validation are summarised on the platform, connected to aforementioned services, with curated job logs highlighting potential warnings and errors.

Successful productions are run on behalf of the user.
Production configurations and the provenance of files are permanently tracked in the \lhcb{\dirac} Bookkeeping database. Users can tag multiple analysis productions indicating that they belong to the same measurement; these tags are used by the {\tt apd}~\cite{apd} tool to allow easy data retrieval as described in Section \ref{sec:apd}. This enables high-quality analysis preservation, as well as efficient data management and archival. The storage usage of {\tt nTuple}s is closely monitored, split by both the physics working group and individual analyses using {\tt apd} tags.

\subsection{Analysis Productions Performance in Run~3}

\begin{figure}
    \centering
    \includegraphics[width=0.8\textwidth]{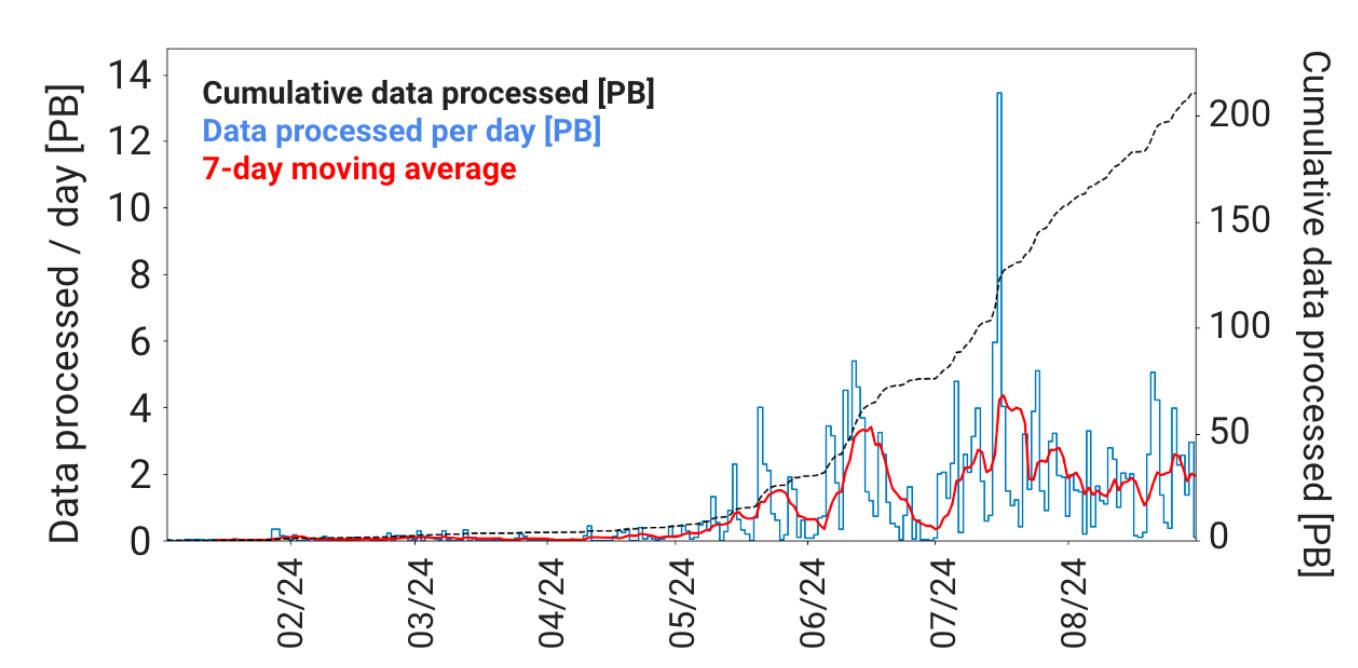}
    \caption{Data processed per day and cumulative data processed by Analysis Productions in 2024.}
    \label{fig:APplot}
\end{figure}

At the time of writing, over 2000 analysis productions have been submitted by analysts on data and simulation. Analysis Productions can be setup to remain ``live" such that they automatically process the data as it arrives out of concurrent Sprucing productions; this means analysts can access their data {\tt nTuples} within days of the data being recorded. Analysis Productions resource uptake follows that of the Sprucing and at its peak in 2024 Analysis Productions were able to process 14\pbytes of data in a single day as shown in Figure~\ref{fig:APplot}. 

\subsection{The {\tt {apd}} Tool and Analysis Preservation}
\label{sec:apd}

The {\tt apd} tool provides a simple programming interface to the Analysis Productions system. It is a Python package published to standard repositories (such as {\tt PyPI}~\cite{pypi} and {\tt conda-forge}~\cite{conda_forge_community_2015_4774216}) that allows the look-up of the physical file names of files created in the context of an Analysis Production, filtering them according to associated metadata (tags). Some of these tags are automatically created by the system (\eg, the data type), but analysts can also define custom tags that indicate the dataset properties or its intended use. Using {\tt apd}, analysts can avoid keeping lists of physical file names that have time-limited validity and instead declare input data with information that represents its properties and intended use (\eg, the analysis name, the working group it is attached to, the data-taking year, the event type in case of simulated data, etc.).

Calls to {\tt apd} within an analysis code base allow clear access to the provenance of the data. To further improve this, an interface to the workflow management system {\tt Snakemake}~\cite{snakemake} was also designed and implemented; {\tt Snakemake} has seen significant adoption in \lhcb analyses. The integration of {\tt apd} into {\tt Snakemake} allows tracking of dependencies between the various data artifacts produced by the analysis up to the original files created by the Analysis Production.

The ease of use is crucial to help with the adoption of any new system. The {\tt apd} tool is installed in the default \lhcb analysis environment which is available through {\tt lb-conda}.  The {\tt lb-conda} tool defines \lhcb Conda wrapper scripts~\cite{lbconda} that provide access to Conda versions on CERN's {\tt cvmfs}~\cite{blomer2011cvmfs}. The {\tt apd} tool has few dependencies and can be installed on any machine independently of the other \lhcb software applications if necessary. Using {\tt{apd}} from Python scripts requires minimal effort from analysts, and also provides the functionality to cache metadata and files locally, to improve performance, or reproduce part of the analysis locally.

Historically LHCb has seen issues with the preservation of analyses due to the evolving nature of computing. This means that even if the analysis and software is well preserved, the ability to connect to external systems to access the data eventually stops working. This can be due to security developments --- for example, the deprecation of the {\tt TLS 1.0} security protocol~\cite{tls} or Certificate Authority changes --- or modernisation of storage systems such as the migration to {\tt EOS}~\cite{eos} at CERN.
The {\tt apd} tool solves this by abstracting the network connectivity through a common interface meaning, in the long term, it is then possible to substitute the service {\tt apd} uses to lookup files with a local {\tt HTTP}~\cite{http} server or even file URLs. Data access with {\tt apd} nominally uses {\tt XRootD}~\cite{xrootd} however this can be transparently changed to return paths to copies on local {\tt POSIX} storage.

Combining the use of {\tt apd} to search for \lhcb Analysis Production data files, {\tt lb-conda} to ensure the reproducibility of the analysis environments and {\tt Snakemake} to define the analysis workflows ensures comprehensive and reliable analysis preservation. The ease of use of this system has led to significant uptake from \lhcb analysts.


\section{The \davinci Application}
\label{sec:davinci}

Analysis Productions uses the offline analysis software project \davinci to create {\tt nTuples} for further high-level analysis. The \davinci application is built on the \gaudi~\cite{gaudi} framework software package for processing High Energy Physics event data. In Run~3 \davinci employs the purpose-built {\tt FunTuple} component to facilitate the storage of event data in {\tt ROOT} format, optimising it for subsequent offline analysis~\cite{Mathad:2023zky}.  These {\tt nTuples} may contain information about the selected signal candidates and the decay products such as particle kinematics, particle identification hypotheses, vertex fit qualities, etc. Within \davinci one can run the \decaytreefitter algorithm~\cite{Hulsbergen:2005pu} on the signal candidates to constrain the particle momenta to point to particular production vertices and constrain the composite particle masses to improve the resolution of the reconstructed objects.

The {\tt nTuples} may also include information about additional particles saved by the selective persistency of the \hlttwo and Sprucing lines. This information may be used to create ``isolation'' information concerning the signal candidate (\ie related to nearby tracks in the event that may be associated with the signal), or for flavour tagging~\cite{prouve2024fastinclusiveflavourtagging}. As the selections in \hlttwo and Sprucing are not exhaustive for every possible resonance, one can also use the extra persisted particles to reconstruct excited states for spectroscopy purposes (for example, adding a track to a \Dz candidate to form a \Dstarp or $D^{\ast\ast+}$).

The \davinci application is built on top of \moore and the rest of the \lhcb software stack that is used for \hlttwo and Sprucing. Therefore, \davinci shares the same algorithms and tools as HLT and Sprucing, namely the {\sc ThOr}~\cite{Thor} based selection and combinatorial functions.

At the time of writing, \davinci writes tuples of {\tt ROOT} columnar \textsc{TTree} objects. In the immediate future, this {\tt nTuple} writing will be updated for thread-safe writing with the \root \textsc{TTreeWriter}. Subsequently, the output type will be changed to the new {\tt RNTuple}~\cite{rntuple}. Doing so will bring significant performance enhancements --- particularly with regard to memory usage and I/O speed --- and significant file size reductions. This is vital to future-proof \davinci, ensuring its usability as data sample sizes grow rapidly towards \lhc Runs 4 and 5.



\section{\lhcb Offline Data Processing and Analysis in \lhcb Upgrade II}

The HL-LHC (LHC Run 4 onwards) is due to begin data collection in 2030. Whereas the general purpose LHC detectors --- ATLAS and CMS --- will take data at the increased luminosity already in Run 4, \lhcb will instead further increase its instantaneous luminosity by a factor five from Run 5 onwards (scheduled for 2036), after the completion of the \lhcb Upgrade II. With the resulting instantaneous luminosity of $1.5\times 10^{34}$ cm$^{-2}$s$^{-1}$, \lhcb aims to achieve a total integrated luminosity of $\approx 300\invfb$ over the lifetime of the HL-LHC.

In Run 5 analysts can naively expect a factor-of-five increase in their {\tt nTuple} data volume. If we maintain current workflows, with analysts still performing significant data processing and filtering steps privately, the {\tt nTuple} data volumes will become unmanageable. Alongside wider file-format and data retrieval R\&D, \lhcb's strategy is to move as much of the data processing and filtering that is currently done by analysts into centralised productions run on WLCG resources, achieved by further exploiting the highly successful Analysis Productions system for more steps in analyses. This is known as the {\tt extended AP model}.  In this model Analysis Productions will perform analysis steps including:

\begin{itemize}
    \item Application of rectangular selection cuts.
    \item Evaluation of derived quantities with subsequent cuts.
     \item Machine Learning model application and inference.
    \item Calibration (\eg, particle identification and tracking efficiencies) routines reducing the number of variables required in final {\tt nTuples}.
    \item Binning of variables for histogram based analyses.
\end{itemize}

Generally, these steps are trivially parallelisable and are well suited to grid productions. WLCG resources are experiment agnostic and can run any code/tool that can be packaged to {\tt cvmfs} or deployed via a container. Portable analysis environments thus allow APs to run any application required for the above steps. In particular, the \lhcb~{\tt lb-conda} tool that provides access to conda environments in {\tt cvmfs} offers a comprehensive (versioned) default environment with all common HEP software packages, as well as the \lhcb calibration tools. Alongside this, analysts can create customised environments specifically for their analysis, which are also versioned, enabling full analysis environment preservation.

In moving these steps to APs, analysts must be able to prototype algorithms and cuts on a subset of the data and train any machine learning models. Analysis Productions allow users to request a prescaled dataset, randomly selecting files across a data-taking period, to prototype their analysis. Once finalised, an analysis production can run over the full dataset applying established algorithms, models and cuts. This workflow is well suited to potential future data storage and retrieval methods that reduce the data stored on disk at any one time through performative data recall from tape.

In the current analysis model, the bespoke submission routines and authentications required by the different distributed resources that analysts use frustrates workflows and collaboration. In the {\tt extended AP model}, analysts will leave the WLCG much later in the analysis process, grid submission routines will be handled for the user by the AP system and users will require only a single, familiar authentication step. The resulting derived datasets are registered in \lhcb bookkeeping with data provenance ensured and are accessible to all \lhcb analysts. The WLCG also benefits from being accessible to the entire LHC virtual organisation, balanced with fair-use policies.

To reduce the data reading load on the WLCG sites, APs will be grouped into so-called ``Analysis Trains" whereby physics working groups combine APs running over the same input data, meaning only a single read of the dataset is required. Re-streaming the Spruced data can also alleviate the I/O load. The analysis train model is also well suited to potential future data storage and retrieval methods that opt for performative data recall from tape.

\lhcb's {\tt extended AP model} is highly dependent on the direction of WLCG over the next decade; \lhcb is helping to guide this direction alongside the needs of the ATLAS and CMS collaborations in Run 4.

\section{Summary}

\lhcb has overhauled its offline data processing and analysis model to ensure full and efficient exploitation of data in Run 3 and beyond; specifically the Sprucing offline data processing stage and centralised Analysis Productions are crucial developments. In Run 5 \lhcb is to increase its instantaneous luminosity by another factor of 5. To manage the increase in data volume of analyst-level data {\tt nTuples}, \lhcb will employ the {\tt extended AP model} through which Analysis Productions will progressively perform all aspects of parallelisable analyst-level data processing and filtering, exploiting the shared resources of WLCG and aligning with potential future data retrieval methods.

\clearpage
\section*{Acknowledgements}

\noindent

The authors express their gratitude to all \lhcb collaborators who have enabled the successful development and uptake of the Sprucing and Analysis Productions. In particular we wish to thank the \lhcb production managers for running the many, many campaigns, the \lhcb data managers for handling our data, and the LHCb Computing project for maintaining and developing the core software upon which we rely.

A.R.W. is supported by UK Research and Innovation under grant \texttt{\#MR/Y01166X/1}. The work of NG is partially supported by the US National Science Foundation through award \texttt{PHY-2411665}.

\addcontentsline{toc}{section}{References}
\bibliographystyle{LHCb}
\bibliography{main,standard,LHCb-PAPER}

\ifx\mcitethebibliography\mciteundefinedmacro
\PackageError{LHCb.bst}{mciteplus.sty has not been loaded}
{This bibstyle requires the use of the mciteplus package.}\fi
\providecommand{\href}[2]{#2}
\begin{mcitethebibliography}{10}
\mciteSetBstSublistMode{n}
\mciteSetBstMaxWidthForm{subitem}{\alph{mcitesubitemcount})}
\mciteSetBstSublistLabelBeginEnd{\mcitemaxwidthsubitemform\space}
{\relax}{\relax}

\bibitem{Alves:2008zz}
LHCb collaboration, A.~A. Alves~Jr.\ {\em et~al.}, \ifthenelse{\boolean{articletitles}}{\emph{{The \lhcb detector at the LHC}}, }{}\href{https://doi.org/10.1088/1748-0221/3/08/S08005}{JINST \textbf{3} (2008) S08005}\relax
\mciteBstWouldAddEndPuncttrue
\mciteSetBstMidEndSepPunct{\mcitedefaultmidpunct}
{\mcitedefaultendpunct}{\mcitedefaultseppunct}\relax
\EndOfBibitem
\bibitem{LHCb:upgradeone}
LHCb collaboration, R.~Aaij {\em et~al.}, \ifthenelse{\boolean{articletitles}}{\emph{{The LHCb Upgrade I}}, }{}\href{https://doi.org/10.1088/1748-0221/19/05/P05065}{JINST \textbf{19} (2024) P05065}, \href{http://arxiv.org/abs/2305.10515}{{\normalfont\ttfamily arXiv:2305.10515}}\relax
\mciteBstWouldAddEndPuncttrue
\mciteSetBstMidEndSepPunct{\mcitedefaultmidpunct}
{\mcitedefaultendpunct}{\mcitedefaultseppunct}\relax
\EndOfBibitem
\bibitem{LHCbCollaboration:2319756}
{LHCb Collaboration},  \ifthenelse{\boolean{articletitles}}{\emph{{Computing Model of the Upgrade LHCb experiment}}}{}, \href{https://cds.cern.ch/record/2319756}{CERN-LHCC-2018-014, LHCb-TDR-018}, CERN, Geneva, 2018\relax
\mciteBstWouldAddEndPuncttrue
\mciteSetBstMidEndSepPunct{\mcitedefaultmidpunct}
{\mcitedefaultendpunct}{\mcitedefaultseppunct}\relax
\EndOfBibitem
\bibitem{Tsaregorodtsev:2010zz}
A.~Tsaregorodtsev {\em et~al.}, \ifthenelse{\boolean{articletitles}}{\emph{{DIRAC3: The new generation of the LHCb grid software}}, }{}\href{https://doi.org/10.1088/1742-6596/219/6/062029}{J.\ Phys.\ Conf.\ Ser.\  \textbf{219} (2010) 062029}\relax
\mciteBstWouldAddEndPuncttrue
\mciteSetBstMidEndSepPunct{\mcitedefaultmidpunct}
{\mcitedefaultendpunct}{\mcitedefaultseppunct}\relax
\EndOfBibitem
\bibitem{LHCb-FIGURE-2020-016}
LHCb collaboration, {LHCb Collaboration},  \ifthenelse{\boolean{articletitles}}{\emph{{RTA and DPA dataflow diagrams for Run 1, Run 2, and the upgraded LHCb detector}}}{}, \href{https://cds.cern.ch/record/2730181}{LHCb-FIGURE-2020-016}, 2020\relax
\mciteBstWouldAddEndPuncttrue
\mciteSetBstMidEndSepPunct{\mcitedefaultmidpunct}
{\mcitedefaultendpunct}{\mcitedefaultseppunct}\relax
\EndOfBibitem
\bibitem{mdf}
M.~Frank, \ifthenelse{\boolean{articletitles}}{\emph{{Online RAW Data Format}}, }{} \url{https://edms.cern.ch/ui/file/784588/1/Online_Raw_Data_Format.pdf"}, 2022\relax
\mciteBstWouldAddEndPuncttrue
\mciteSetBstMidEndSepPunct{\mcitedefaultmidpunct}
{\mcitedefaultendpunct}{\mcitedefaultseppunct}\relax
\EndOfBibitem
\bibitem{rfc}
Y.~Collet and M.~Kucherawy, \ifthenelse{\boolean{articletitles}}{\emph{{Zstandard Compression and the ``application/zstd" Media Type}}, }{} RFC 8878, 2021.
\newblock doi:~\href{https://doi.org/10.17487/RFC8878}{10.17487/RFC8878}\relax
\mciteBstWouldAddEndPuncttrue
\mciteSetBstMidEndSepPunct{\mcitedefaultmidpunct}
{\mcitedefaultendpunct}{\mcitedefaultseppunct}\relax
\EndOfBibitem
\bibitem{bird2011wlcg}
I.~Bird, \ifthenelse{\boolean{articletitles}}{\emph{Computing for the {Large Hadron Collider}}, }{}\href{https://doi.org/10.1146/annurev-nucl-102010-130059}{Annual Review of Nuclear and Particle Science \textbf{61} (2011) 99}\relax
\mciteBstWouldAddEndPuncttrue
\mciteSetBstMidEndSepPunct{\mcitedefaultmidpunct}
{\mcitedefaultendpunct}{\mcitedefaultseppunct}\relax
\EndOfBibitem
\bibitem{Moore}
{LHCb Collaboration}, \ifthenelse{\boolean{articletitles}}{\emph{{Moore application GitLab repository}}, }{}
\newblock \url{https://gitlab.cern.ch/lhcb/Moore}\relax
\mciteBstWouldAddEndPuncttrue
\mciteSetBstMidEndSepPunct{\mcitedefaultmidpunct}
{\mcitedefaultendpunct}{\mcitedefaultseppunct}\relax
\EndOfBibitem
\bibitem{DaVinci}
{LHCb Collaboration}, \ifthenelse{\boolean{articletitles}}{\emph{{DaVinci application GitLab repository}}, }{}
\newblock \url{https://gitlab.cern.ch/lhcb/DaVinci/}\relax
\mciteBstWouldAddEndPuncttrue
\mciteSetBstMidEndSepPunct{\mcitedefaultmidpunct}
{\mcitedefaultendpunct}{\mcitedefaultseppunct}\relax
\EndOfBibitem
\bibitem{Thor}
{LHCb Collaboration}, \ifthenelse{\boolean{articletitles}}{\emph{{ThOr functors}}, }{}
\newblock \url{https://lhcbdoc.web.cern.ch/lhcbdoc/moore/master/selection/thor_functors.html}\relax
\mciteBstWouldAddEndPuncttrue
\mciteSetBstMidEndSepPunct{\mcitedefaultmidpunct}
{\mcitedefaultendpunct}{\mcitedefaultseppunct}\relax
\EndOfBibitem
\bibitem{Aaij:2147693}
R.~Aaij {\em et~al.}, \ifthenelse{\boolean{articletitles}}{\emph{{Tesla : an application for real-time data analysis in High Energy Physics}}, }{}Comput.\ Phys.\ Commun.\  \textbf{208} (2016) \href{http://arxiv.org/abs/1604.05596}{{\normalfont\ttfamily arXiv:1604.05596}}\relax
\mciteBstWouldAddEndPuncttrue
\mciteSetBstMidEndSepPunct{\mcitedefaultmidpunct}
{\mcitedefaultendpunct}{\mcitedefaultseppunct}\relax
\EndOfBibitem
\bibitem{git}
L.~Torvalds and J.~Hamano, \ifthenelse{\boolean{articletitles}}{\emph{Git distributed version control system}, }{} \url{http://git-scm.com}, 2010.
\newblock Accessed: 2024-06-04\relax
\mciteBstWouldAddEndPuncttrue
\mciteSetBstMidEndSepPunct{\mcitedefaultmidpunct}
{\mcitedefaultendpunct}{\mcitedefaultseppunct}\relax
\EndOfBibitem
\bibitem{gitlab}
{GitLab Inc.\ }, \ifthenelse{\boolean{articletitles}}{\emph{Gitlab}, }{} \url{https://about.gitlab.com}, 2011.
\newblock Accessed: 2024-06-04\relax
\mciteBstWouldAddEndPuncttrue
\mciteSetBstMidEndSepPunct{\mcitedefaultmidpunct}
{\mcitedefaultendpunct}{\mcitedefaultseppunct}\relax
\EndOfBibitem
\bibitem{arXiv:2503.19582}
L.~Grazette {\em et~al.}, \ifthenelse{\boolean{articletitles}}{\emph{{A Comprehensive Bandwidth Testing Framework for the LHCb Upgrade T Trigger System}}, }{}\href{http://arxiv.org/abs/2503.19582}{{\normalfont\ttfamily arXiv:2503.19582}}\relax
\mciteBstWouldAddEndPuncttrue
\mciteSetBstMidEndSepPunct{\mcitedefaultmidpunct}
{\mcitedefaultendpunct}{\mcitedefaultseppunct}\relax
\EndOfBibitem
\bibitem{LHCb-TDR-011}
LHCb collaboration, \ifthenelse{\boolean{articletitles}}{\emph{{LHCb computing: Technical Design Report}}, }{} \href{http://cdsweb.cern.ch/search?p=CERN-LHCC-2005-019&f=reportnumber&action_search=Search&c=LHCb} {CERN-LHCC-2005-019}, 2005\relax
\mciteBstWouldAddEndPuncttrue
\mciteSetBstMidEndSepPunct{\mcitedefaultmidpunct}
{\mcitedefaultendpunct}{\mcitedefaultseppunct}\relax
\EndOfBibitem
\bibitem{ROOT}
R.~Brun and F.~Rademakers, \ifthenelse{\boolean{articletitles}}{\emph{{ROOT} — an object oriented data analysis framework}, }{}\href{https://doi.org/10.1016/S0168-9002(97)00048-X}{Nucl.\ Instrum.\ Meth.\ A \textbf{389} (1997) 81}\relax
\mciteBstWouldAddEndPuncttrue
\mciteSetBstMidEndSepPunct{\mcitedefaultmidpunct}
{\mcitedefaultendpunct}{\mcitedefaultseppunct}\relax
\EndOfBibitem
\bibitem{jsroot}
ROOT, \ifthenelse{\boolean{articletitles}}{\emph{{JavaScript ROOT}}, }{}
\newblock \url{https://github.com/root-project/jsroot}\relax
\mciteBstWouldAddEndPuncttrue
\mciteSetBstMidEndSepPunct{\mcitedefaultmidpunct}
{\mcitedefaultendpunct}{\mcitedefaultseppunct}\relax
\EndOfBibitem
\bibitem{lumiplotcitation}
{LHCb Collaboration}, \ifthenelse{\boolean{articletitles}}{\emph{{{End of successful proton-proton collision data taking period}}}, }{}
\newblock https://lhcb-outreach.web.cern.ch/2024/10/18/end-of-successful-proton-proton-collision-data-taking-period/\relax
\mciteBstWouldAddEndPuncttrue
\mciteSetBstMidEndSepPunct{\mcitedefaultmidpunct}
{\mcitedefaultendpunct}{\mcitedefaultseppunct}\relax
\EndOfBibitem
\bibitem{bkk}
{Mathe, Z.\ },  \ifthenelse{\boolean{articletitles}}{\emph{{Feicim: A browser and analysis tool for distributed data in particle physics.}}}{}, \href{http://cds.cern.ch/record/1491175}{CERN-THESIS-2012-156}, 2012\relax
\mciteBstWouldAddEndPuncttrue
\mciteSetBstMidEndSepPunct{\mcitedefaultmidpunct}
{\mcitedefaultendpunct}{\mcitedefaultseppunct}\relax
\EndOfBibitem
\bibitem{prmon}
G.~A. Stewart and A.~S. Mete, \ifthenelse{\boolean{articletitles}}{\emph{prmon: process monitor}, }{} 2024.
\newblock doi:~\href{https://doi.org/10.5281/ZENODO.11400398}{10.5281/ZENODO.11400398}\relax
\mciteBstWouldAddEndPuncttrue
\mciteSetBstMidEndSepPunct{\mcitedefaultmidpunct}
{\mcitedefaultendpunct}{\mcitedefaultseppunct}\relax
\EndOfBibitem
\bibitem{apd}
C.~Burr, B.~Couturier, and R.~O’Neil, \ifthenelse{\boolean{articletitles}}{\emph{{Facilitating the preservation of LHCb Analyses with APD}}, }{}\href{https://doi.org/10.1051/epjconf/202429508008}{EPJ Web of Conferences \textbf{295} (2024) 08008}\relax
\mciteBstWouldAddEndPuncttrue
\mciteSetBstMidEndSepPunct{\mcitedefaultmidpunct}
{\mcitedefaultendpunct}{\mcitedefaultseppunct}\relax
\EndOfBibitem
\bibitem{pypi}
{Python community}, \ifthenelse{\boolean{articletitles}}{\emph{{Python Package Index PyPI}}, }{}
\newblock \url{https://pypi.org/}\relax
\mciteBstWouldAddEndPuncttrue
\mciteSetBstMidEndSepPunct{\mcitedefaultmidpunct}
{\mcitedefaultendpunct}{\mcitedefaultseppunct}\relax
\EndOfBibitem
\bibitem{conda_forge_community_2015_4774216}
{Conda-forge community}, \ifthenelse{\boolean{articletitles}}{\emph{{The conda-forge Project: Community-based Software Distribution}}, }{} 2015.
\newblock doi:~\href{https://doi.org/10.5281/zenodo.4774216}{10.5281/zenodo.4774216}\relax
\mciteBstWouldAddEndPuncttrue
\mciteSetBstMidEndSepPunct{\mcitedefaultmidpunct}
{\mcitedefaultendpunct}{\mcitedefaultseppunct}\relax
\EndOfBibitem
\bibitem{snakemake}
J.~Köster and S.~Rahmann, \ifthenelse{\boolean{articletitles}}{\emph{Snakemake---a scalable bioinformatics workflow engine}, }{}\href{https://doi.org/10.1093/bioinformatics/bts480}{Bioinformatics \textbf{28} (2012) 2520}\relax
\mciteBstWouldAddEndPuncttrue
\mciteSetBstMidEndSepPunct{\mcitedefaultmidpunct}
{\mcitedefaultendpunct}{\mcitedefaultseppunct}\relax
\EndOfBibitem
\bibitem{lbconda}
{LHCb Collaboration}, \ifthenelse{\boolean{articletitles}}{\emph{{LbCondaWrappers GitLab repository}}, }{}
\newblock \url{https://gitlab.cern.ch/lhcb-core/lbcondawrappers}\relax
\mciteBstWouldAddEndPuncttrue
\mciteSetBstMidEndSepPunct{\mcitedefaultmidpunct}
{\mcitedefaultendpunct}{\mcitedefaultseppunct}\relax
\EndOfBibitem
\bibitem{blomer2011cvmfs}
J.~Blomer {\em et~al.}, \ifthenelse{\boolean{articletitles}}{\emph{{CernVM File System}}, }{}\href{https://doi.org/10.1088/1742-6596/331/4/042003}{Journal of Physics: Conference Series \textbf{331} (2011) 042003}\relax
\mciteBstWouldAddEndPuncttrue
\mciteSetBstMidEndSepPunct{\mcitedefaultmidpunct}
{\mcitedefaultendpunct}{\mcitedefaultseppunct}\relax
\EndOfBibitem
\bibitem{tls}
C.~Allen and T.~Dierks, \ifthenelse{\boolean{articletitles}}{\emph{{The TLS Protocol Version 1.0}}, }{}RFC 2246 (1999), Online; accessed 2025-06-23\relax
\mciteBstWouldAddEndPuncttrue
\mciteSetBstMidEndSepPunct{\mcitedefaultmidpunct}
{\mcitedefaultendpunct}{\mcitedefaultseppunct}\relax
\EndOfBibitem
\bibitem{eos}
A.~J. Peters and L.~Janyst, \ifthenelse{\boolean{articletitles}}{\emph{Exabyte scale storage at cern}, }{}\href{https://doi.org/10.1088/1742-6596/331/5/052015}{Journal of Physics: Conference Series \textbf{331} (2011) 052015}\relax
\mciteBstWouldAddEndPuncttrue
\mciteSetBstMidEndSepPunct{\mcitedefaultmidpunct}
{\mcitedefaultendpunct}{\mcitedefaultseppunct}\relax
\EndOfBibitem
\bibitem{http}
R.~Fielding {\em et~al.}, RFC \ifthenelse{\boolean{articletitles}}{\emph{{Hypertext Transfer Protocol -- HTTP/1.1}}}{}, \href{https://www.rfc-editor.org/info/rfc2616}{2616}, Internet Engineering Task Force (IETF), 1999.
\newblock Obsoleted by RFCs 7230, 7231, 7232, 7233, 7234, 7235, doi:~\href{https://doi.org/10.17487/RFC2616}{10.17487/RFC2616}\relax
\mciteBstWouldAddEndPuncttrue
\mciteSetBstMidEndSepPunct{\mcitedefaultmidpunct}
{\mcitedefaultendpunct}{\mcitedefaultseppunct}\relax
\EndOfBibitem
\bibitem{xrootd}
{XRootD}, \ifthenelse{\boolean{articletitles}}{\emph{{XRootD project}}, }{}
\newblock \url{http://www.xrootd.org/}\relax
\mciteBstWouldAddEndPuncttrue
\mciteSetBstMidEndSepPunct{\mcitedefaultmidpunct}
{\mcitedefaultendpunct}{\mcitedefaultseppunct}\relax
\EndOfBibitem
\bibitem{gaudi}
G.~Barrand {\em et~al.}, \ifthenelse{\boolean{articletitles}}{\emph{Gaudi — a software architecture and framework for building hep data processing applications}, }{}\href{https://doi.org/https://doi.org/10.1016/S0010-4655(01)00254-5}{Computer Physics Communications \textbf{140} (2001) 45}, CHEP2000\relax
\mciteBstWouldAddEndPuncttrue
\mciteSetBstMidEndSepPunct{\mcitedefaultmidpunct}
{\mcitedefaultendpunct}{\mcitedefaultseppunct}\relax
\EndOfBibitem
\bibitem{Mathad:2023zky}
A.~Mathad {\em et~al.}, \ifthenelse{\boolean{articletitles}}{\emph{{FunTuple: A New N-tuple Component for Offline Data Processing at the LHCb Experiment}}, }{}\href{https://doi.org/10.1007/s41781-024-00116-1}{Comput.\ Softw.\ Big Sci.\  \textbf{8} (2024) 6}, \href{http://arxiv.org/abs/2310.02433}{{\normalfont\ttfamily arXiv:2310.02433}}\relax
\mciteBstWouldAddEndPuncttrue
\mciteSetBstMidEndSepPunct{\mcitedefaultmidpunct}
{\mcitedefaultendpunct}{\mcitedefaultseppunct}\relax
\EndOfBibitem
\bibitem{Hulsbergen:2005pu}
W.~D. Hulsbergen, \ifthenelse{\boolean{articletitles}}{\emph{{Decay chain fitting with a Kalman filter}}, }{}\href{https://doi.org/10.1016/j.nima.2005.06.078}{Nucl.\ Instrum.\ Meth.\  \textbf{A552} (2005) 566}, \href{http://arxiv.org/abs/physics/0503191}{{\normalfont\ttfamily arXiv:physics/0503191}}\relax
\mciteBstWouldAddEndPuncttrue
\mciteSetBstMidEndSepPunct{\mcitedefaultmidpunct}
{\mcitedefaultendpunct}{\mcitedefaultseppunct}\relax
\EndOfBibitem
\bibitem{prouve2024fastinclusiveflavourtagging}
C.~Prouve, N.~Nolte, and C.~Hasse, \ifthenelse{\boolean{articletitles}}{\emph{{Fast Inclusive Flavour Tagging at LHCb}}, }{}\href{http://arxiv.org/abs/2404.14145}{{\normalfont\ttfamily arXiv:2404.14145}}\relax
\mciteBstWouldAddEndPuncttrue
\mciteSetBstMidEndSepPunct{\mcitedefaultmidpunct}
{\mcitedefaultendpunct}{\mcitedefaultseppunct}\relax
\EndOfBibitem
\bibitem{rntuple}
J.~Blomer {\em et~al.}, \ifthenelse{\boolean{articletitles}}{\emph{{ROOT’s RNTuple I/O Subsystem: The Path to Production}}, }{}\href{https://doi.org/10.1051/epjconf/202429506020}{EPJ Web of Conf.\  \textbf{295} (2024) 06020}\relax
\mciteBstWouldAddEndPuncttrue
\mciteSetBstMidEndSepPunct{\mcitedefaultmidpunct}
{\mcitedefaultendpunct}{\mcitedefaultseppunct}\relax
\EndOfBibitem
\end{mcitethebibliography}
 
\end{document}